\documentclass[fleqn,12pt,twoside]{article}
\usepackage{espcrc1}
\usepackage{epsfig}  

\usepackage{graphicx}
\usepackage[figuresright]{rotating}

\newcommand{\AmS}{{\protect\the\textfont2
  A\kern-.1667em\lower.5ex\hbox{M}\kern-.125emS}}

\title{Charm and leptons}

\author{P.~Crochet\address[]{Laboratoire de Physique Corpusculaire\\
    CNRS/IN2P3 and Universit\'e Blaise Pascal\\
    F-63000 Clermont-Ferrand, France}
    \thanks{crochet@clermont.in2p3.fr}}
       
\begin{document}

\maketitle

\begin{abstract}
The present knowledge on charm and leptons is reviewed including the topics 
discussed at this conference and the progress made since the last Quark Matter
conference. Special emphasis is placed on $J/\psi$ production at the SPS which
is one of the highlights in the field.
\end{abstract}

\section{Introduction}

Observables related to charm and leptons are of great relevance in heavy ion 
collisions considering the significant information that they have revealed 
over the last years.
The low mass dielectron yield measured by the CERES collaboration has
provided evidence for in-medium effects on the $\rho$ meson.
The $J/\psi$ suppression pattern observed by the NA50 collaboration is 
believed to be an important signature of the QGP.
The intermediate mass dimuon excess in A-A reactions at SPS has 
been first attributed to an increase of the open charm cross-section.
At the beginning of this year, the first measurement of the open charm 
cross-section in heavy ion collisions has been carried-out at RHIC by the 
PHENIX collaboration.
On the other hand, the actual mechanism for charmonium production in heavy 
ion reactions is a subject of intensive current debate.
Indeed, the new scenarios which have been recently proposed differ totally 
from the standard approach in their predictions. 
At RHIC they imply a significantly larger $J/\psi$ yield than 
that resulting from the standard suppression scenario.
This makes the first $J/\psi$ measurements from PHENIX of crucial importance.

\section{Low mass dileptons at SPS}

Studies of low mass dielectrons at the SPS continue to attract a great 
interest since the observation of an excess beyond the expected sources in S-U
reactions at $200~A{\rm GeV}$~\cite{ceresfirst} and in 
Pb-Au reactions at $158~A{\rm GeV}$~\cite{ceressecond}.
The enhancement is located in the invariant mass range 
$0.2<m<0.7~{\rm GeV/c}^2$.
It is found to be more pronounced at low $p_{\rm t}$ and to increase with
centrality.
The final analysis of the $40~A{\rm GeV}$ Pb-Au data presented at this 
conference~\cite{hannes02} confirms the preliminary results reported at the 
last Quark Matter conference~\cite{harry}:
the data taken at $40~A{\rm GeV}$ exhibit an even larger enhancement 
($5.1\pm1.3({\rm stat})\pm1.0({\rm syst})$) than at $158~A{\rm GeV}$ 
($2.9\pm0.3({\rm stat})\pm0.6({\rm syst})$).
This is probably due to the maximum baryonic density reached 
in Pb-Au collisions at the lower beam energy.
Note that, within errors, the measured dielectron yield is, in the considered 
invariant mass range, similar at both beam energies such that the larger 
excess observed at $40~A{\rm GeV}$ actually results from a lower magnitude of 
the sum of known hadronic decay sources.
The enhancement can be well accounted for by considering reduction in mass
or broadening in width of the $\rho$ meson in the hot and dense medium 
(see~\cite{rappreport} for a review and~\cite{rappupdate} for a recent update).
Unfortunately, the present experimental resolution does not allow to 
disentangle the two scenarios (this might be improved with the analysis of the
data taken in 2000).
Another approach, recently proposed, is based on dileptons from a simple 
parameterization of a thermal source~\cite{galldileptons}. 
This approach is very successful in describing also the intermediate mass 
dilepton yield and the direct photons at SPS (see below).
The derived space-time averaged temperature is $170(145)~{\rm MeV}$ at 
$158(40)~A{\rm GeV}$. 
It is interesting to note that the temperature at $158~A{\rm GeV}$ coincide 
with the expected temperature of the phase transition.
Its maximum is located at $210~{\rm MeV}$ suggesting that the system
originates from the deconfined region~\cite{galldileptons}.

\section{Intermediate mass dileptons at SPS}

In the intermediate mass range (IMR, $1.5<m<2.5~{\rm GeV/c}^2$), like in the 
case of the low mass region, an excess of dileptons is seen compared to
known sources in A-A reactions.
This has been observed by the HELIOS/3~\cite{heliosocharm} and 
NA50~\cite{na50ocharm} collaborations.
An updated analysis was presented at the last Quark Matter 
conference~\cite{cap01}.
The enhancement is found to increase linearly with ${\rm N}_{\rm part}$ and
reaches a factor $\sim2.1$ for the most central Pb-Pb events.
Surprisingly, the kinematical distributions of the dimuon excess are 
compatible with those expected from open charm decay~\cite{na50ocharm}.
These findings have triggered a lot of theoretical activities with various 
interpretations: 
open charm enhancement~\cite{oc-wong}, 
$D$ meson final state rescattering~\cite{oc-lin},
secondary hadron production~\cite{oc-li},
secondary Drell-Yan production~\cite{oc-spieles},
$\eta_c$ production~\cite{oc-anishetty},
thermal dilepton production~\cite{oc-rapp},
secondary meson-baryon interactions~\cite{oc-cassing},
in-medium Drell-Yan production~\cite{oc-qiu}.
Some of the proposed scenarios fail in reproducing the kinematical 
distributions of the excess, some fail in its centrality dependence 
and some underpredict its magnitude.
The most successful description of the data has been achieved by means of 
thermal dileptons~\cite{oc-rapp} (see also~\cite{oc-gale}).
Indeed, with a common thermal source one can simultaneously reproduce the low 
mass dielectron yield from CERES, the IMR dimuons from NA50 and HELIOS/3 
as well as the direct photons from WA98~\cite{galldileptons}.
However, it should be mentioned that a recent investigation of the centrality 
dependence of the IMR dimuon excess leads to a source temperature which is 
significantly larger than that estimated from central events 
only~\cite{oc-kampfer}.
This could leave room for a partial contribution from one or some of the other 
effects listed above.

One should finally note that, from the experimental point of view, this kind 
of analysis requires a thorough understanding of the combinatorial background.
This is because, in contrast to narrow resonances, the present broad signal 
does not clearly stick-out of the background.
In particular, special attention should be paid to situations with a
production asymmetry between positive and negative leptons and a large signal 
which will lead to an incorrect background subtraction and therefore 
to a biased extracted signal~\cite{marek}.

\section{Open charm at RHIC}

The first (indirect) measurement of the open charm cross-section in heavy ion 
collisions has been carried-out recently by the PHENIX collaboration via 
analysis of single electron spectra in Au-Au at 
$\sqrt{s}=130~{\rm GeV}$~\cite{phenixPRL}.
After subtracting all known sources from the total distribution the remaining
electron spectra are, in the explored $p_{\rm t}$ range, in good agreement 
with electron spectra from open charm decay as predicted using PYTHIA 
(tuned to reproduce existing hadron-nucleon data at lower beam energies and 
then extrapolated to Au-Au at RHIC).
The derived integrated cross-section is in good agreement with PYTHIA and 
consistent with pQCD NLO calculations.
Therefore, in contrast to the SPS IMR dimuon data, there is no need for charm 
enhancement in order to explain the RHIC single electron spectra.
However, because of the present large systematical errors, one cannot firmly 
exclude in-medium effects on charm production at RHIC. 
It was actually pointed-out at this conference that the small differences at 
high $p_{\rm t}$ between the minimum-bias spectrum and the central spectrum 
could be explained by considering energy loss of charm 
quarks~\cite{kampferqm02}.

The analysis was repeated for the $200~{\rm GeV}$ data~\cite{ralf02}.
Here, the determination of the background was done in a completely different 
way, namely by using a photon converter instead of the cocktail technique 
which was used for the $130~{\rm GeV}$ data.
Again, the data agree well with the expectations from PYTHIA. 
This holds true for several centrality bins such that, here again, not much 
room is left for large in-medium effects.
However, it is clear that the safest way to make a definite statement about 
in-medium effects on open charm at RHIC is to compare the Au-Au data to the 
p-p data. 
This is hopefully going to be done soon~\cite{ralf02}.

Another interesting perspective, with the coming high luminosity 
runs, is the independent measurement of the open charm cross-section from the 
dielectron continuum above the $\phi$ meson.
All these measurements could provide precious information on open bottom 
by looking at high $p_{\rm t}$ and high invariant mass. 
They would be greatly improved by means of high resolution vertex
detectors owing to the large $c\tau$ of open charm and open bottom hadrons. 
Note finally that non-direct measurements of open charm (bottom) from 
semi-leptonic decays bring limited information only. 
Direct measurements of open charm hadrons remain an important goal of
the heavy quark physics program at RHIC.

\section{Charmonium at SPS}

After many years of intensive experimental and theoretical investigations, 
charmonium production at SPS remains one of the hottest topic in the 
field (see~\cite{satzram} for reviews).
This is because the observed behaviour of $J/\psi$ is believed to exhibit one 
of the most pronounced deviation with respect to pure conventional hadronic 
picture.

The so-called ``$J/\psi$ anomalous suppression'' has been observed by the NA50 
collaboration with the data collected in 1996 and 1998.
It manifests itself by a departure from the ``normal'' nuclear absorption when
looking at the centrality dependence of the $(J/\psi)/({\rm DY})$ ratio 
in Pb-Pb reactions~\cite{na50jpsiET,na50jpsiZDC}.
The ratio is obtained by means of either the standard method or the 
minimum-bias method. 
The centrality of the reactions is measured with the neutral transverse energy
(${\rm E}_{\rm t}$) and the Zero Degree Calorimeter energy
(${\rm E}_{\rm ZDC}$).
A $\sim20\%$ drop is observed at ${\rm E}_{\rm t}\sim40~{\rm GeV}$, followed 
by a decrease at ${\rm E}_{\rm t}\sim100~{\rm GeV}$ (Tab.~\ref{tab1}).


The NA50 collaboration interprets this pattern as an evidence for 
deconfinement. 
In this scenario, the drop results from the melting of 
$\chi_c$\footnote{The fraction of $J/\psi$ from $\chi_c$ decay is $\sim30\%$ 
  in ${\rm p}$-$\bar{\rm p}$ at $\sqrt{s}=1.8~{\rm TeV}$~\cite{cdf} and
  $\sim32\%$ in p-A at $\sqrt{s}=41.6~{\rm GeV}$~\cite{herab}.} 
and the decrease from the melting of $J/\psi$.
This can be checked~\cite{gerschel} by comparing the energy densities derived 
from the data (Tab.~\ref{tab1}) with the expected charmonia dissociation 
temperatures from recent studies based on the heavy quark potential from 
LQCD~\cite{digal}. 
The data agree with the predictions within $\sim20\%$.
This is a fairly good agreement considering the uncertainties in the way the 
energy density is derived from the data.
Note however that these studies assume the $\chi_c$ decay branching ratio into
$J/\psi$ in heavy ion reactions to be the same as in p-p.

\vspace*{-0.5cm}
\begin{table}[htb]
  \caption{Location of the drop and the decrease with
    the transverse energy (${\rm E}_{\rm t}$), 
    the Zero Degree Calorimeter energy (${\rm E}_{\rm ZDC}$),
    the path length through matter (${\rm L}$),
    the impact parameter (${\rm b}$), 
    the number of participant nucleons (${\rm N}_{\rm part}$)
    and the energy density ($\epsilon$).
    The values are from the data collected in 1996 and 
    1998~\cite{na50jpsiET,na50jpsiZDC}.}
\label{tab1}
\newcommand{\m}{\hphantom{$-$}}
\newcommand{\cc}[1]{\multicolumn{1}{c}{#1}}
\renewcommand{\tabcolsep}{0.92pc} 
\renewcommand{\arraystretch}{1.2} 
\begin{tabular}{@{}lllllll}
\hline
  & \cc{${\rm E}_{\rm t}~{\rm (GeV)}$} & \cc{${\rm E}_{\rm ZDC}~{\rm (TeV)}$} 
  & \cc{${\rm L}~{\rm (fm)}$}  & \cc{${\rm b}~{\rm (fm)}$} 
  & \cc{${\rm N}_{\rm part}$} & \cc{$\epsilon~{\rm (GeV/fm}^3{\rm )}$}\\
\hline
drop        & \m41  & \m25 & \m8   & \m8 & \m122 & \m2.3 \\
decrease    & \m100 & \m9  & \m9.3 & \m3 & \m334 & \m3.1 \\
\hline
\end{tabular}\\[2pt]
\end{table}
\vspace*{-0.7cm}

\subsection{$(J/\psi)/({\rm DY})$ in central Pb-Pb}

The interpretation of the decrease as due to the melting of $J/\psi$ is not 
straightforward in a dynamical picture.
Indeed, QGP models which assume two sharp thresholds corresponding to the 
successive meltings of $\chi_c$ and $J/\psi$ overpredict the data in the 
decrease (i.e. beyond ${\rm E}_{\rm t}=100~{\rm GeV}$)~\cite{nardi}. 
In fact, it was realized that at ${\rm E}_{\rm t}=100~{\rm GeV}$ the 
${\rm E}_{\rm t}$ distribution exhibits the typical knee beyond which there
is no further sensitivity to centrality and fluctuations set 
in~\cite{blaizotPRL2}.
By taking into account these fluctuations, a perfect fit of the data is 
achieved over all the centrality range with either two thresholds or 
one gradual threshold~\cite{blaizotPRL2}. 
In other words, in the framework of QGP models, the decrease seen in the 
data cannot be non-ambiguously associated to the melting of $J/\psi$.

Furthermore, it has recently been pointed-out that the decrease may hide 
another effect which has strictly nothing to do with the QGP but which is 
related to the trigger of the NA50 spectrometer~\cite{cap2nddrop}.
An event which fulfills the $J/\psi$ trigger conditions should have, in 
average, a smaller ${\rm E}_{\rm t}$ than a minimum-bias event because a 
fraction of ${\rm E}_{\rm t}$ is taken away by the $J/\psi$.
Therefore, in the minimum-bias analysis, the $(J/\psi)/({\rm DY})$ ratio is
biased because the $J/\psi$ yield is measured (under trigger conditions) 
whereas the Drell-Yan yield is calculated (without trigger conditions).
It has been shown in~\cite{cap2nddrop} that this effect leads to an extra 
depletion of the $(J/\psi)/({\rm DY})|_{\rm min-bias}$ ratio in the 
${\rm E}_{\rm t}$ range of the decrease.
Consequently, with ${\rm E}_{\rm t}$ fluctuations and ${\rm E}_{\rm t}$ loss 
effects, the comover model provides a very good description of the data up to 
the most central events~\cite{cap2nddrop}.
This also holds true for other models based on different
scenarios~\cite{kosJPG28,loic022,chaud} and would probably also hold true for 
former models which assume or not the formation of the QGP~\cite{nonQGPmodel}.
In this context the decrease cannot be considered any longer as a valid 
point to disprove a model which would exhibit a saturation of the 
$(J/\psi)/({\rm DY})$ ratio at high ${\rm E}_{\rm t}$.
Note that the trigger ${\rm E}_{\rm t}$ loss effect can, in principle, 
be confirmed or disproved by taking Drell-Yan data at large ${\rm E}_{\rm t}$
and/or by measuring at RHIC the $J/\psi$ yield at the energy density 
corresponding to that of the decrease at the SPS.

\subsection{$(J/\psi)/({\rm DY})$ in peripheral Pb-Pb}

At the time of the last Quark Matter conference, the $(J/\psi)/({\rm DY})$
data at low ${\rm E}_{\rm t}$ were subject to different interpretations.
Indeed, as it can be seen from Fig.~8 of~\cite{bord01}, at low 
${\rm E}_{\rm t}$, some models underpredict the data, some 
overpredict the data and some go through the data.
On the other hand, the data points below ${\rm E}_{\rm t}=30~{\rm GeV}$
are located above the nuclear absorption curve which means that they show
less absorption than in p-A and S-U systems.
The same feature is observed at large ${\rm E}_{\rm ZDC}$~\cite{na50jpsiZDC}
(i.e. for the same centrality class).
This is difficult to understand in the QGP picture since the suppression 
should be ``normal'' below the (first) threshold.
However, it was believed that, for these peripheral events, the data could 
be contaminated by beam-air interactions.
In order to clarify the situation new data were taken in 2000 with the target
placed in the vacuum to minimize off-target interactions. 
The software was also improved with new methods for rejection of pile-up events
and identification of the interaction point.
In parallel, more accurate measurements of p-A data were made in order to 
estimate with better precision the $J/\psi$ nuclear absorption cross-section.
The updated value is 
$\sigma^{\rm abs}_{J/\psi}=4.4\pm0.5~{\rm mb}$~\cite{cort02}
(the former was 
$\sigma^{\rm abs}_{J/\psi}=6.4\pm0.8~{\rm mb}$~\cite{na50PLB466}).
The new $(J/\psi)/({\rm DY})|_{\rm standard}$ data have been presented at 
this conference~\cite{rame02}.
They are consistent with the previously published data for 
${\rm E}_{\rm t}>30~{\rm GeV}$. 
On the contrary, for ${\rm E}_{\rm t}<30~{\rm GeV}$, they lie significantly 
below the former data and are compatible with the (new) nuclear absorption 
curve.
Consequently, models going through the previous data points at low 
${\rm E}_{\rm t}$ should strongly deviate from the new data in this 
${\rm E}_{\rm t}$ range. 
In particular, this is the case for the comover 
model~\cite{capfluct,cap2nddrop}.

\vspace*{0.3cm}

Conclusion: 
It has been shown at this conference~\cite{rame02} that the published 
peripheral $(J/\psi)/({\rm DY})$ data suffer from a 
systematical bias due to beam-air interactions 
(a $\sim25\%$ effect at ${\rm E}_{\rm t}=15~{\rm GeV}$). 
It has been argued in~\cite{cap2nddrop} that the central 
$(J/\psi)/({\rm DY})|_{\rm min-bias}$ data might be affected by a systematical
bias due to trigger ${\rm E}_{\rm t}$ loss (a $\sim20\%$ effect at 
${\rm E}_{\rm t}=120~{\rm GeV}$). 
Very recently, it has been claimed that there might be systematical 
inconsistencies between the ${\rm E}_{\rm t}$ data and the ${\rm E}_{\rm ZDC}$
data for semi-central events~\cite{capetzdc}.
As interpretations are often being drawn at a few percent level, 
it is extremely important that the data are carefully checked for 
systematical effects.
In addition, more stringent tests of the models could certainly be achieved
i) by studying simultaneously the centrality dependence of yields together 
with ratios and
ii) by extending investigations to all available observables.
The latter should include, in particular, the transverse momentum dependence 
of the suppression discussed below.

\subsection{Transverse momentum dependence of the suppression}

The $p_{\rm t}$ dependence of the $J/\psi$ suppression was early considered 
to be the golden observable to prove the existence of the QGP.
Traditionally the data are presented in terms of the centrality dependence of 
the $<p_{\rm t}^2>$ of $J/\psi$ in order to provide evidence for the 
$p_{\rm t}$ broadening effect due to initial state parton scattering which is 
expected in absence of QGP.
In a deconfined medium, it was first believed that one should observe 
additional $p_{\rm t}$ broadening with increasing centrality~\cite{ptbroad}. 
This is because, due to the resonance formation time and screening effects, 
only high $p_{\rm t}$ $J/\psi$ could escape the medium whereas low 
$p_{\rm t}$ $J/\psi$ would melt.
Later on, it was claimed~\cite{kharzeev97} that an opposite effect should
actually happen, namely $<p_{\rm t}^2>$ flattening or even decreasing.
The reason is that high $p_{\rm t}$ $J/\psi$ are those which travel through 
the largest amount of matter and therefore are those which come from the 
center of the reaction. 
Since the local density is the highest in the central region these 
high $p_{\rm t}$ $J/\psi$ should be the most QGP suppressed.
On the other hand, comovers are expected to produce a weak 
$<p_{\rm t}^2>$ flattening~\cite{gavin96,armesto99}.

The NA50 Pb-Pb data~\cite{na50PBpt} are shown in Fig.~1 
(see~\cite{topi02} for the data taken in 2000 
and~\cite{na50SUpt} for the S-U data at $200~A{\rm GeV}$).
They show a smooth increase of $<p_{\rm t}^2>$ with centrality and 
a saturation when approaching mid-central reactions.
It is astounding to note that the data do not show any significant sign for 
onsets, steps or drops and it would be remarkable that the two QGP effects 
mentioned above would exactly cancel each other. 
Even more striking is the outcome of the comparison with the theory, since
the only model which significantly misses the data is the QGP model
(first noticed in~\cite{drapier}).
This means that something is inconsistent when looking simultaneously at 
$(J/\psi)/({\rm DY})$ and at $<p_{\rm t}^2>$ of $J/\psi$.
It is therefore of crucial importance to further investigate these data 
carefully.
To this respect, it is also important to stress that one would surely learn
much more from the spectra themselves instead of limiting the study to mean 
values\footnote{As illustrated in~\cite{e877}, the fact that a model gets 
the mean of a distribution right does not necessarily mean that the full 
distribution is understood.}.

\vspace*{-0.5cm}
\begin{figure}[hbt]
\label{myfig}
\begin{minipage}[t]{10.cm}
  {\epsfig{file=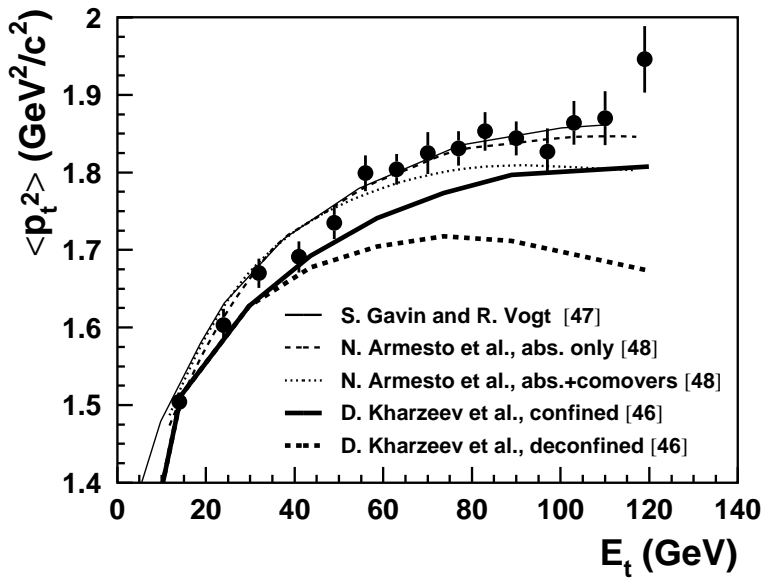, width=1.0\linewidth}}
\end{minipage}
\hspace{\fill}
\begin{minipage}[t]{5.6cm}
\vspace*{-8.5cm}
\caption{$<p_{\rm t}^2>$ of $J/\psi$ as a function of the transverse 
  energy 
  in Pb-Pb reactions at $158~A{\rm GeV}$.
  The data (dots)~\cite{na50PBpt} are compared to various model
  predictions (curves) with references reported on the figure. 
  Some of the calculations were performed at a beam energy of 
  $200~A{\rm GeV}$ instead of $158~A{\rm GeV}$.
  In order to account for this, the $<p_{\rm t}^2>$ values 
  have been rescaled according to~\cite{drapier}.}
\end{minipage}
\end{figure}
\vspace*{-0.5cm}


Another interesting aspect of the centrality dependence of the $J/\psi$ 
$p_{\rm t}$ spectra has been presented at this conference~\cite{topi02}.
The $J/\psi$ inverse slope parameters for p-A, S-U and non-central Pb-Pb 
reactions fall on a common straight line when plotted as a function of the 
energy density. 
However, for the most-central Pb-Pb events one observes a flattening 
of the apparent temperatures. 
This could indicate a change in the expansion of the system.

\subsection{Non-direct charmonium production}

Charmonium production from statistical hadronization has been extensively
discussed in a parallel session at this conference. 
This production mechanism is based on a twofold observation~\cite{pbmPLB00}. 
First, because of cross-section considerations, the treatment of charmed 
hadrons in a standard thermal model cannot be easily justified.
Secondly, in central Pb-Pb reactions at SPS the measured 
$\psi^\prime/(J/\psi)$ ratio is $\sim4\%$ as predicted by the thermal model.
This led to a new scenario for charmonium production~\cite{pbmPLB00}: 
i) all $c\bar{c}$ pairs are produced in the early stage via hard scattering;
ii) all charm hadrons are produced statistically at the hadronization.
The underlying picture is that the charm hadrons either do not form in the 
early stages or are fully suppressed in the QGP.
This scenario can be implemented in the framework of a thermal model by using 
the formalism described in~\cite{pbmPLB00,gorPLB509,anton02} where one 
includes a charm enhancement factor in order to match the multiplicity of 
directly produced $c\bar{c}$ pairs. 
Since the open charm cross-section has not yet been measured at SPS, it
is taken from the predictions of pQCD.

This approach has been studied in detail in the two last
years~\cite{pbmPLB00,gorPLB509,anton02,kosPLB531,kosJPG28,gorJPG28,loic02,loic022}. 
As shown in~\cite{anton02} the measured $J/\psi$ yield versus 
${\rm N}_{\rm part}$ in Pb-Pb is qualitatively consistent with the centrality 
dependence of the nuclear overlap function.
In order to reproduce the yields, one needs to enhance the charm cross-section
in the model with respect to the predictions from pQCD.
This extra charm enhancement amounts to a factor $\sim2$. 
This is surprisingly similar to the open charm-like enhancement needed in 
order to account for the excess of the IMR dimuon yield observed in central 
Pb-Pb events at SPS~\cite{cap01}.
Another realization of the statistical hadronization model also provides a 
very good fit of the ${\rm E}_{\rm t}$ dependence of the $(J/\psi)/({\rm DY})$
ratio, albeit with a larger enhancement factor~\cite{kosPLB531,kosJPG28}.

An additional experimental indication for statistical hadronization of 
charmonium at SPS has been put forward in~\cite{bugaevSPS} 
(see~\cite{bugaevRHIC} for the model predictions at RHIC):
The $m_{\rm t}$ spectra of $\Omega^\pm$ from WA97 and $J/\psi$ and 
$\psi^\prime$ from NA50 can be simultaneously described in a hydrodynamical
picture of hadronizing QGP with ${\rm T}={\rm T}_{\rm c}=170~{\rm MeV}$ and 
$\bar{v}=0.2$ ($\bar{v}$ is the average transverse flow velocity).
This supports the idea of charmonium formation and kinetic freeze-out at the 
hadronization.

The statistical hadronization model can also be used to predict charmonium 
yield at RHIC~\cite{anton02,gorJPG28}.
At RHIC, depending on the beam energy, the total multiplicity of (directly 
produced) $c\bar{c}$ pairs is expected to be smaller than one in peripheral 
reactions and larger than one in central reactions.
Because the statistical hadronization process is strongly correlated to the
multiplicity of the $c\bar{c}$ pairs, one expects the final 
$(J/\psi)/({\rm N}_{c\bar{c}})$ ratio to increase or decrease versus 
centrality depending on the beam energy~\cite{gorJPG28}.

The $J/\psi$ statistical hadronization process has been recently incorporated
in a modified approach~\cite{loic02,loic022}. 
This approach combines i) directly produced $J/\psi$ followed by nuclear
absorption and QGP/HG dissociation and ii) statistically produced $J/\psi$ as 
in pure statistical models but without the extra charm enhancement factor. 
This leads to a good agreement with the NA50 $(J/\psi)/({\rm DY})$ data.
The agreement is particularly good at the level of the first drop which arises,
in this model, from the superposition of direct and statistically 
produced $J/\psi$.
Another interesting aspect from this model is the excitation function
of the total $J/\psi$ yield which is expected to be dominated by direct 
$J/\psi$ at SPS and by statistical $J/\psi$ at RHIC, with 
a minimum at $\sqrt{s}\sim50~{\rm GeV}$.
It is important to note that whereas pure statistical models reproduce the 
measured $\psi^\prime/(J/\psi)$ ratio, this approach significantly overpredicts
it. 
This leaves room for additional effects~\cite{loic02}.

Non-direct charmonium states can be 
produced not only at the hadronization but also in the QGP by means of the 
so-called kinetic recombination process~\cite{thewsALL}.
This mechanism foresees even larger $J/\psi$ yields at RHIC than the 
statistical hadronization process.
Note that all models will face stringent tests with the upcoming data from 
NA60 at SPS and PHENIX at RHIC.

To conclude on this topic, it is interesting to extrapolate from charmonium 
production at RHIC to bottomonium production at LHC.
Indeed, i) the key parameter for statistical production of charmonium is the 
number of $c\bar{c}$ pairs and ii) the expected multiplicity of $b\bar{b}$ 
pairs at LHC is roughly equal to the expected multiplicity of $c\bar{c}$ 
pairs at RHIC. 
Therefore, if statistical production of charmonium is confirmed at RHIC, one 
could expect the same mechanism to occur at LHC for bottomonium states.

\section{Charmonium at RHIC}

The first charmonium measurements from the PHENIX collaboration are certainly 
among the most exciting results presented at this conference~\cite{fraw02}.
The $J/\psi$ signal has been extracted in p-p at $\sqrt{s}=200~{\rm GeV}$ both
in the dielectron channel and in the dimuon channel. 
Although the present statistics is very limited ($\sim24(36)$ counts in the 
dielectron(dimuon) channel), the signal is clearly seen.
Thanks to the capability of the PHENIX detector to measure simultaneously
the two channels, the first rapidity distribution of $J/\psi$ in p-p
at $\sqrt{s}=200~{\rm GeV}$ has been established. 
The $J/\psi$ signal has also been extracted in Au-Au reactions 
in the dielectron channel. 
The statistics, even more limited than in p-p, does not allow to draw at 
present solid conclusions on a possible $J/\psi$ suppression or enhancement.
However, large suppresion factors as well as large enhancement factors seem
to be not likely.
This will be improved soon since only half of the collected data from the 
$200~{\rm GeV}$ run has been analysed so far.
In the next two years the statistics will be enhanced by a factor $\sim75$ 
for Au-Au and $\sim100$ for p-p. 
Also of great interest will be the coming charmonium measurements in d-A.

\section{Next steps}

Apart from the forthcoming measurements at RHIC and the analysis of remaining 
SPS data, new generation experiments are presently being completed or designed.

The HADES experiment~\cite{hades} at GSI will soon shed light on the DLS 
low mass dielectron puzzle~\cite{dls}.
The NA60 experiment~\cite{na60} at SPS will continue in line 
with the investigations made by NA50 and, in particular, measure the open 
charm cross-section and the $\psi^\prime$.
The ALICE experiment~\cite{paolo} will bring essential information regarding 
charm and leptons at LHC.
Looking even forward into the future, the CBM experiment~\cite{cbm}, at the 
GSI future facility, will allow a new beam energy domain to be explored with 
simultaneous measurements of low mass dielectrons, charmonia and open charm.
 
\section*{Acknowledgements}

Stimulating discussions with A.~Andronic, R.~Averbeck, P.~Braun-Munzinger, 
M.I. Gorenstein, J.Y.~Ollitrault, L.~Ramello and J.~Stachel are gratefully 
acknowledged.
Many thanks to J.P.~Cussonneau, C.~Finck, R.~Granier de Cassagnac and 
F.~Staley for technical support, to the speakers who provided their slides 
prior to their talk and to the organizers of the conference.

\end{document}